 \documentclass[final,5p,times,authoryear]{elsarticle}
 
\usepackage{amssymb}
\usepackage{lipsum}
\usepackage[T1]{fontenc} % if needed
\usepackage{booktabs} % To thicken table lines
\usepackage{ dsfont }
\usepackage{float}% better control of floating figures and tables

\usepackage{calrsfs}
\DeclareMathAlphabet{\pazocal}{OMS}{zplm}{m}{n}
\usepackage{xcolor}
\usepackage{pict2e}
\usepackage[percent]{overpic}
\usepackage{ upgreek }
\usepackage{xfrac}
\usepackage{abraces}% http://ctan.org/pkg/abraces
\setcitestyle{numbers,square,comma,sort&compress}

\usepackage{braket}
\usepackage{stmaryrd}
\usepackage{amsmath,empheq}
\usepackage{multirow}

\journal{Physics Letters B}

\begin{document}

\begin{frontmatter}



\title{Retuning radio astronomy for axion dark matter with neutron stars}


\author[a,b,c]{Javier De Miguel\corref{cor1}}
\address[a]{Instituto de Astrofísica de Canarias, E-38200 La Laguna, Tenerife, Spain}
\address[b]{Departamento de Astrofísica, Universidad de La Laguna, E-38206 La Laguna, Tenerife, Spain}
\address[c]{The Institute of Physical and Chemical Research (RIKEN), Center for Advanced Photonics, 519-1399 Aramaki-Aoba, Aoba-ku, Sendai, Miyagi 980-0845, Japan}
\cortext[cor1]{jdemiguel@iac.es}

\begin{abstract}
A model is constructed to predict the emission originating from axion-to-photon conversion in the strongly magnetized ultrarelativistic plasma of neutron stars. The acceleration and multiplicity of the charges  are observed to shift the axion-induced spectral feature with respect to previous expectations. The frequency range of interest widens accordingly, and heavier dark matter axions may resonate in magnetospheric splits giving rise to detectable radio signals that could extend into the millimeter band. Ultimately, this work follows an affirmative answer to the question of whether neutron stars can give rise to any detectable high-frequency spectral feature that would allow us to probe axion dark matter of masses up to about a millielectronvolt. SGR 1745--2900 emerges as a particularly promising astrophysical laboratory for probing high-frequency axion dark matter.
\end{abstract}




\end{frontmatter}

%\tableofcontents



\section{Introduction}
The presence of a intriguing non-luminous substance was inferred from astronomical observations almost a century ago by Zwicky \cite{ 1933AcHPh...6..110Z}, and later by Rubin \& Ford \cite{1970ApJ...159..379R}—see \cite{2018RvMP...90d5002B} for a review. The Quantum Chromodynamics (QCD) axion is a long-postulated pseudo-scalar boson that arises in a plausible explanation for the charge and parity symmetry problem in the strong interaction \cite{PhysRevLett.38.1440, PhysRevLett.40.223, PhysRevLett.40.279}. In addition, the axion is a compelling candidate to non-relativistic, or `cold,' dark matter (DM) in a broad mass, $m_{a}$, \textit{versus} coupling strength to photons, $g_{a\gamma}$, parameter space \cite{ABBOTT1983133, DINE1983137, PRESKILL1983127}. The relevant interaction reads
\begin{equation}
\pazocal{L}_{a\gamma}= -\frac{1}{4}g_{a\gamma}a F_{\!\mu \nu} \tilde{F}^{\mu \nu} \;,
\label{Eq.1}
\end{equation}
with $a$ as the axion field, $F^{\mu \nu}$ being the photon field; dual fields are denoted with a \textit{tilde}. From a classic perspective, the term $\pazocal{L}_{a\gamma}=g_{a\gamma}a\mathrm{E}\cdot\mathrm{B}$ is yield from Eq. \ref{Eq.1},
where E is the electromagnetic wave and B the external, static magnetic field; where the role of the polarization in the mixing of axion and two photons via the Primakoff effect \cite{Primakoff:1951iae} is manifested through the dot product.

The potential of neutron stars (NSs), evolved compact objects exhibiting ultra strong magnetic fields, in the role of benchmark astrophysical laboratories to probe Galactic axion DM, was noted four decades ago, including Refs. \cite{Iwamoto:1984ir, PhysRevD.37.1237, Morris:1984iz, Yoshimura:1987ma}. These pioneering works relied on the NS model developed by Goldreich \& Julian (GJ), in which a corotational, charge-separated plasma, with a balanced occupation of electron-proton pairs, fills the light-cylinder, assuming a perfect conductivity and homogeneity in the dipole field limit \cite{1969ApJ...157..869G}. More recent works have continued the same approach, shaping the emission signal originating from the resonant conversion of axion-like Galactic DM falling on NSs adopting a GJ-like density profile \cite{Lai:2006af, PhysRevLett.121.241102, Huang:2018lxq, Leroy:2019ghm, Safdi:2018oeu, PhysRevD.102.023504, Witte:2021arp, Battye:2021xvt}. Unfortunately, despite the efforts made, no trace of axion has been found in the centimeter-wavelength spectra of NSs \cite{Darling:2020plz, Darling:2020uyo, PhysRevLett.125.171301, PhysRevLett.125.171301, Foster:2022fxn}, which is associated with the range of axion masses that are resonant in a GJ-type stellar magnetosphere—i.e., below a few dozen $\upmu$eV. 

However, the absence of DM evidence to date does not necessarily imply that the search for the axion with NSs has ended unsuccessfully. Moreover, it is known that primary charges multiply in the magnetosphere of neutron stars by electromagnetic cascade \cite{1975ApJ...196...51R, 1979ApJ...231..854A, 1997A&A...322..846K,Rea:2008zs, Timokhin:2015dua, Timokhin:2018vdn, Lyutikov:2007xw, 2007ApJ...658.1177D, Olmi:2016avl, 10.1111/j.1365-2966.2010.17449.x}, to which state-of-the-art axion conversion models have not devoted their attention because the adopted GJ model neglects pair cascades. In this concern, this manuscript takes care of a star model beyond the GJ approach that may alter the paradigm of the search for the spectral feature originating from the resonant conversion of the axion in NSs once a pair multiplicity factor is inserted which modifies the density profile of the magnetosphere along with the relativistic acceleration of the charges that shift the resonance frequency. In this respect, since the mass and coupling of the QCD axion are barely known, it is crucial to cover the widest possible frequency range with the maximum achievable sensitivity. Therefore, probing NSs also at higher frequencies in search for the axion could be an engaging counterpart to a variety of detection experiments and phenomenological considerations—e.g., see \cite{PhysRevD.105.035022, PhysRevD.98.103015, Marsh_2017, PhysRevLett.118.011103, 2014PhRvL.113s1302A, Straniero:2015nvc, 2022JCAP...10..096D, 2022JCAP...02..035D, REGIS2021136075, 2021Natur.590..238B, EHRET2010149, PhysRevD.88.075014, 2016EPJC...76...24D, PhysRevD.92.092002,PhysRevLett.59.839, PhysRevD.42.1297, PhysRevLett.104.041301,PhysRevD.97.092001,PhysRevLett.128.241805,doi:10.1063/5.0098783,2021Natur.590..238B, CAST:2017uph, CAST:2020rlf, PhysRevLett.121.261302, PhysRevD.99.101101, PhysRevD.103.102004, MCALLISTER201767, doi:10.1126/sciadv.abq3765}.

The rest of this article is organized as follows. In Sec. \ref{II}, an axion-to-photon conversion model that accommodates both the multiplicity of pairs and the relativistic velocities of the accelerated charges is built-up. The application of the model to pulsars and their populations in the inner region of our galaxy in search for axion DM are addressed in Sec. \ref{III}. The particular case of SGR 1745--2900 is treated in Sec. \ref{Ap2} to give presence to the case of magnetars, extremely magnetic NSs that may require peculiar modelling. Some relevant conclusions are drawn in Sec. \ref{IV}. Two appendices are annexed that are devoted to: identifying the sources for which the model is transferable—\ref{A1}—; \ref{Ap3} represents an initial step toward a more generalized and precise axion-photon mixing model, capable of simultaneously accounting for acceleration and charge multiplicity while accommodating the propagation of oblique modes across a broader range of sources.

\section{General model}\label{II}
A widespread strategy to deal with the electromagnetic cascades while maintaining a radial density profile is to account for the overdensity of pairs by comparison with the model provided by GJ, which defines the minimum corotational charge density as a function of the angular velocity of the star and the magnetic field, $n_{_{\mathrm{GJ}_e}}= 2\Omega_* B(r)/e$. This allows to define the pair multiplicity factor as the number of electron and positron pairs produced
by each primary particle, $\pazocal{M}=n_{e}/n_{_{\mathrm{GJ}}}$, that
could range between 1 and $>$$10^5$ \cite{DeMiguel:2021pfe,Rea:2008zs, Timokhin:2015dua, Timokhin:2018vdn, Lyutikov:2007xw, 2007ApJ...658.1177D, Olmi:2016avl, 10.1111/j.1365-2966.2010.17449.x}; likely greater in the incipience of giant flares \cite{PhysRevD.106.L041302}. Pair injection and kinematics may allow the magnetosphere to be filled with plasma in excess of the GJ density \cite{Philippov_2014, Philippov_2015, 10.1093/mnras/stv042, Brambilla_2018, refId0Gu, Chen_2020}.

\textit{Tour de force} works provide a theoretical framework to shape the emission line originating from the resonant conversion of axion into photons in the magnetosphere of isolated NSs, including Refs. \cite{PhysRevLett.121.241102, Leroy:2019ghm, PhysRevD.102.023504}. Those models are revisited in this manuscript in order to account for: (i) a charge production factor beyond the boundaries of GJ, and (ii) an (ultra-)relativistic plasma instead of the usual non-relativistic approach. To this end, it is denoted $\hat{m}$ as the magnetic axis in a misaligned rotator, termed $\hat{z}$; the propagation direction, $\hat{r}$; the magnetic angle with respect to the rotation direction, $\theta_m$; with $\theta$ as the polar angle from the rotation axis. The dipole approximation is adopted, at which $B(r)=1/2B_0(R_*/r)^3f_1(t)$, where the latter term gives account on the dependence on time, $t$. The charge density of the magnetospheric plasma is \cite{PhysRevE.57.3399}
\begin{equation}
n_{e}=\frac{2\,\Omega_* B(r)}{e} \, \frac{\pazocal{M}} {\gamma_p} \,,
\label{Eq.2}
\end{equation}

with $\gamma_p$ as the Lorentz factor of the plasma charges, while the characteristic plasma frequency is
\begin{equation}
\omega^2_p= \frac{4\pi\alpha \,n_{e}(r)}{m_e}  \,,
\label{Eq.3}
\end{equation}

with $e$ and $m_e$ as the electron charge and mass, respectively, and $\alpha$ being the fine-structure constant. The gyro-frequency is greater than the photon frequency—and the plasma frequency—, i.e. $\omega_B=e B(r)/m_e \gg \omega$, that allows for the usual simplifications in the dielectric tensor \cite{DeMiguel:2021pfe}. For realistic values of theta $\theta_m$, as $\theta\rightarrow 90^\circ$ Langmuir ordinary (L-O) modes gradually vanish while the ordinary (O) -mode gains weight. In this form, the dispersion relation for O-like modes becomes
\begin{equation}
k^2 \simeq \omega^2-2\omega^2_p \langle \gamma_p ^{-3} \rangle  \,,
\label{Eq.10}
\end{equation}

where $\langle \gamma_p ^{-3} \rangle \sim \langle \gamma_p  \rangle  ^{-1} $  \cite{PhysRevE.57.3399}; the averaging operator being $\langle ...  \rangle_i=\int F_i(p)dp$, with $F_i$ as the distribution function of species $i$. Locally, we have $  \langle \gamma_p  \rangle  \sim \gamma_p$, so Eq. \ref{Eq.10} simplifies to $k\sim | \omega |$ in a relativistic plasma with $\gamma_p\gg 1$. This renders the problem one-dimensional (1D) and analytically treatable at the cost of adding a small error—since $\mathrm{cos}^2\theta\ll1$. Note that for $\pazocal{M}=1=\gamma_p$, the classic GJ approach is recuperated since the approximation $k\sim | \omega |$ also holds in the so-called `weak dispersion' limit \cite{PhysRevLett.121.241102, Leroy:2019ghm, PhysRevD.102.023504}, or for the `high-frequency' limit in O-modes \cite{PhysRevE.57.3399}.

For poorly relativistic axions, the axion-photon oscillation takes place at the point, with a distance from the surface $r_c$, at which the axion and effective photon mass or, equivalently, plasma frequency, resonate, $m_a\sim\omega_p$. Photons are then radiated from the magnetosphere with a frequency $\omega\sim m_a$. The conversion altitude from the surface becomes 
\begin{equation}
\begin{aligned}
r_c\simeq224 \,\mathrm{km} \times f_1^{1/3}(t) \times \frac{R_*}{10 \,\mathrm{km}} \times \\ \left[ \frac{B_0}{10^{14}\,\mathrm{G}}\times \frac{1\,\mathrm{s}}{P}\times\left(\frac{1\,\mathrm{GHz}}{m_a}\right)^2\times \frac{\pazocal{M}_c}{\gamma_c} \right]^{1/3}
\;.
\label{Eq.4}
\end{aligned}
\end{equation}

The velocity of the axion at the conversion point, $v_c$, accumulated by gravitational acceleration, reads $v^2_c\simeq 2GM_*/r_c$; with $M_*$ being the stellar mass; $B_0$ the magnetic field near the stellar surface; $P$ the rotation period; where $G$ is the gravitational constant, while $f_1(t)=|3\,\mathrm{cos}\,\theta\,\hat{m}\cdot\hat{r}-\mathrm{cos}\,\theta_m|$ holds in Eq. \ref{Eq.4}, where $\hat{m}\cdot\hat{r}=\mathrm{cos}\,\theta_m\,\mathrm{cos}\,\theta+\mathrm{sin}\,\theta_m\,\mathrm{sin}\,\theta\,\mathrm{cos}(\Omega_*\,t)$. Typical values for $r_c$ and $v_c$ are a few stellar radii and $c/10$. 
The conversion probability for non-relativistic axions is $
P_{a\gamma}\sim1/2 g^2_{a\gamma} B^2(r_c)\,r_c/{m_a/v_c}$ \cite{PhysRevD.102.023504, Leroy:2019ghm}, where the propagation angle is set  $\theta\rightarrow\pi/2$ causing a negligible error, as mentioned \cite{PhysRevLett.121.241102}. On the other hand, the power emitted per unit of solid angle is $d\pazocal{P}/d\Omega\approx2 P_{a\gamma}\rho_c v_c r^2_c$, where the axion DM density at the conversion region is $\rho_{c}\simeq2/\sqrt{\pi}\rho_{\infty}v_c/v_0$; $\rho_{\infty}$ being the occupation of DM in the vicinity of the star. The flux density at the position of the observer in the line of sight (l.o.s.) is S$_{\nu}=d\pazocal{P}/d\Omega/d^2/\Delta\nu$, $d$ being distance to emitter and $\Delta\nu$ the linewidth. In this regard, in \cite{PhysRevD.97.123001} is claimed that the emission line is Doppler broadened as $\Delta\nu/\nu \sim v_0/c$ due to the radial velocity of the isolated NS. Differently, in \cite{PhysRevLett.121.241102} it is pointed out that the signal is broadened with a leading order term $(v_0/c)^2$ due to energy conservation. Lastly, in \cite{PhysRevD.102.023504, PhysRevLett.125.171301} authors consider the plasma at the conversion point as a spinning mirror causing a broadening that scales as $\Delta\nu/\nu \sim \Omega_* r_c \varepsilon^2/c$, with $\Omega_*=2\pi/P$, which is therefore more pronounced for shorter rotation periods; with $\varepsilon$ being the eccentricity for an oblique rotator at which the intersection of a plane
perpendicular to the rotation axis with the conversion surface
is projected an ellipse. Wider line broadening dilutes the spectral density function. In spite of this, the latter approach is adopted for this article, as it would represent the vaster mechanism in the event that the three widenings competed. Substitutions yield \cite{Darling:2020plz}
\begin{equation}
\begin{aligned}
\mathrm{S}_{\nu}\approx 1.6\times10^{-3} \, \mathrm{\upmu Jy} \times f_2(t) \times \left(\frac{100\,\mathrm{pc}}{d}\right)^2 \times \\ \left(\frac{g_{a\gamma}}{10^{-12}\,\mathrm{GeV^{-1}}}\right)^2  \times \frac{R_*}{10\,\mathrm{km}}\times  \left(\frac{m_a}{1\,\mathrm{GHz}}\right)^{4/3}\times \\ \left(\frac{B_0}{10^{14}\,\mathrm{G}}\right)^{1/3}\times \left(\frac{\Omega_*}{1\,\mathrm{Hz}}\right)^{-8/3}\times  \frac{\rho_{\infty}}{0.4\,\mathrm{GeV cm^{-3}}} \times \\ \frac{M_*}{1\,\mathrm{M_{\odot}}}\times
\frac{200\, \mathrm{km\,s^{-1}}}{v_0}\times \frac{v_c} {c}\times \left(\frac{\pazocal{M}_c}{\gamma_c}\right)^{-3/2} \times \frac{\Omega_c}  {\Omega}
\;,
\label{Eq.7}
\end{aligned}
\end{equation}
which depends on the physical parameters of the star and its density profile at the conversion zone, then on $\pazocal{M}_c$ and the Lorentz factor at the resonance region, $\gamma_c$; the velocity of the star, $v_0$; the spatial configuration, the time, and the occupation and parameter space of the axion. As in \cite{PhysRevLett.121.241102}, I extract the time and polar angle dependence to a $f_2(t)=[3 (\hat{m}\cdot\hat{r})^2+1]\,f^{-4/3}_1(t)$ function in Eq. \ref{Eq.7}, which adopts values of the order of the unity. Eccentricity has been considered close to one. The sensitivity of radio-telescopes in terms of the accessible axion-photon coupling strength is straightforward
\begin{equation}
\begin{aligned}
g_{a\gamma}\gtrsim 2.5\times10^{-11} \, \mathrm{GeV^{-1}} \times \left(\frac{\mathrm{S}_{\nu}}{\mathrm{1\,\upmu Jy}}\right)^{1/2}\times f^{-1/2}_2(t) \times \\ \frac{d}{100\,\mathrm{pc}}\times \left(\frac{10\,\mathrm{km}}{R_*}\right)^{1/2}\times  \left(\frac{\mathrm{1\,GHz}}{m_a}\right)^{2/3}\times \left(\frac{10^{14}\,\mathrm{G}}{B_0}\right)^{1/6}\times \\ \left(\frac{\Omega_*}{\mathrm{1\,Hz}}\right)^{4/3}\times  \left(\frac{0.4\,\mathrm{GeV cm^{-3}}} {\rho_{\infty}}\right)^{1/2}\times  \left(\frac{\mathrm{1\,M_{\odot}}}{M_*}\right)^{1/2}\times \\
\left(\frac{v_0}{200\, \mathrm{km\,s^{-1}}}\right)^{1/2}\times \left(\frac {c}{v_c}\right)^{1/2}\times \left( \frac{\pazocal{M}_c}{\gamma_c}\right)^{3/4} \times \left( \frac  {\Omega} {\Omega_c}  \right)^{1/2}
\;.
\label{Eq.9}
\end{aligned}
\end{equation}

Lastly, the appreciation that vacuum polarization  would drastically mitigate the cross-section of the axion-photon-photon vertex at high frequencies in a highly magnetized star modelable by the GJ profile was noted early on by Raffelt \& Stodolsky \cite{PhysRevD.37.1237}. Nevertheless, from a pair density beyond the frontiers of GJ that emerges on the right hand side of Eq. \ref{Eq.2}, it follows that quantum electrodynamics (QED) corrections to axion electrodynamics \cite{Wilczek:1987mv} due to vacuum birefringence vanish for relatively low frequencies or comparatively large pair multiplicity factors \cite{DeMiguel:2021pfe}. This is the case considered throughout this work.

\section{Pulsars}\label{III}
\begin{figure}[h]\centering
\includegraphics[width=.38\textwidth]{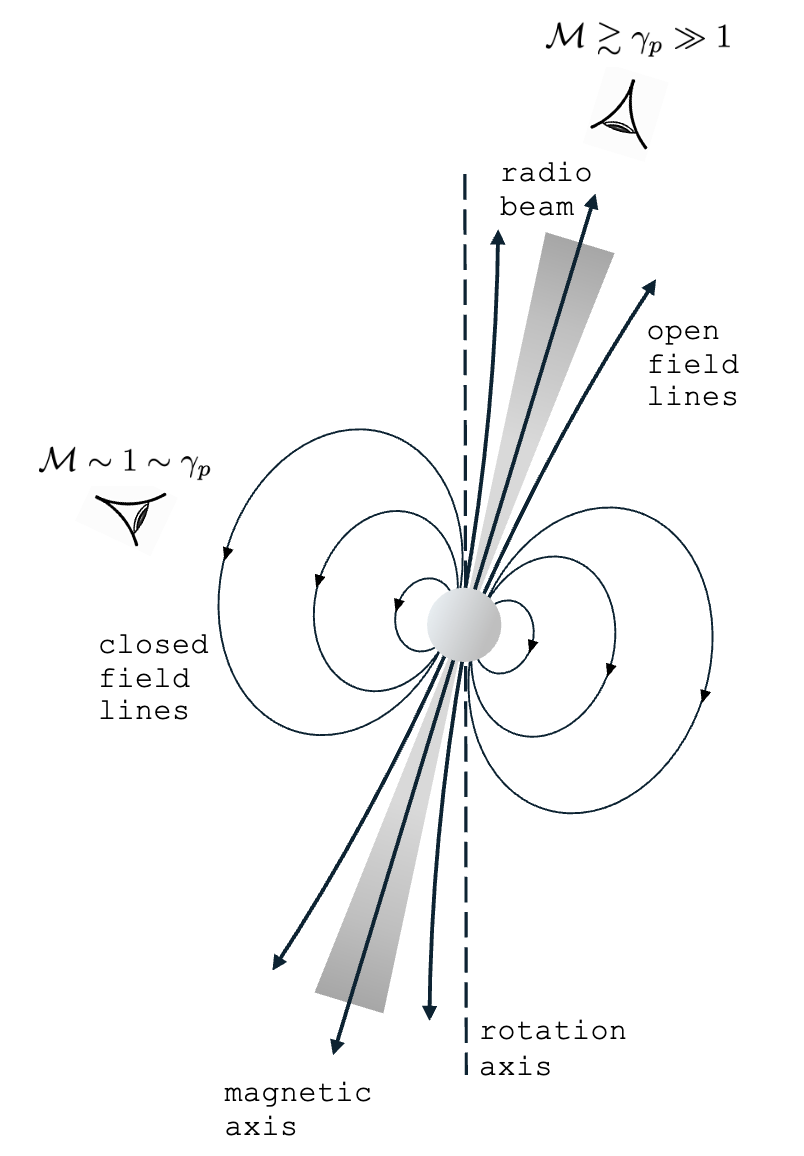}
\caption{A schematic of a pulsar. During data acquisition, an observer along the l.o.s. marked as $\pazocal{M}\sim1\sim \gamma_p$ will observe a region associated with resonant axion-photon conversion within the GJ framework. A telescope aligned with the l.o.s. of $\pazocal{M}\gtrsim \gamma_p\gg1$ will also integrate regions linked to axion conversion into millimeter waves originating near open field lines. Since the emission in the region $\pazocal{M}\gtrsim \gamma_p\gg1$ appears as a pulse along the observer's l.o.s., the steady and pulsed components can be separated using a Fourier transform and analyzed independently. }
\label{fig_R2}
\end{figure}

The classic picture of a pulsating star, or pulsar \cite{1968Natur.217..709H}, is sustained by the creation of pair cascades and particle migration, accelerating the charges to relativistic velocities, giving raise to curvature radiation, synchrotron emission and inverse Compton scattering \cite{1975ApJ...196...51R, 1979ApJ...231..854A, 1997A&A...322..846K, 10.1093/mnras/stv1405}. The magnetosphere is permeated by dense pair plasma, which effectively screens the accelerating electric field across most regions, leaving only a few small zones responsible for most particle acceleration and emission. These regions are referred to as the `polar cap,' extending over a small volume close to the surface which is being traversed by the magnetic axis; the `outer gap,' located at higher altitudes near the light cylinder; and the
`slot gap,' starting near the stellar surface along closed field lines. Inside the polar cap, at the base of the open magnetic field lines, the magnetosphere is expected to contain regions with unscreened electric fields. The pairs generated here have their own dynamics, moving and evolving as they are accelerated along the magnetic field lines. The relativistic pairs can also emit pair-producing photons, initiating an avalanche that continues until photons from the final generation of pairs can no longer produce additional pairs and escape the magnetosphere. The pair plasma exits the magnetosphere along open magnetic field lines, supplying the radiating particles that form the surrounding pulsar wind nebulae—e.g., see \cite{Timokhin:2015dua}.

In this section, we address the open question of whether a high-frequency signal induced by heavier axions resonating in regions of high multiplicity and density exists and is detectable.  \begin{figure}[h]\centering
\includegraphics[width=.5\textwidth]{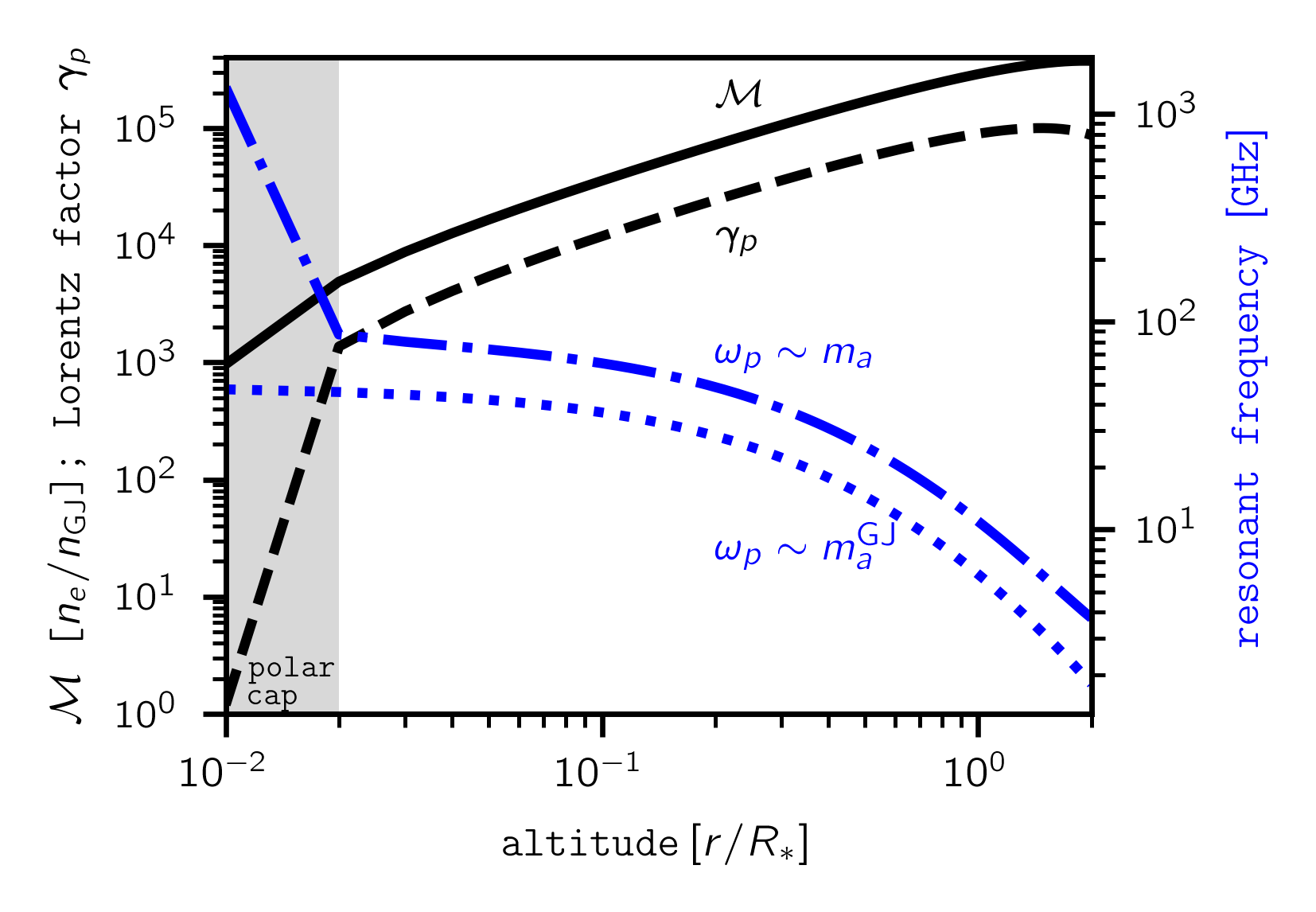}
\caption{Minimal model for a young pulsar as inspired by Timokhin et al. \cite{Timokhin:2015dua, Timokhin:2018vdn} for the range $0.01<r/R_*<2$, with $r$ as the altitude from the star surface. The multiplicity of pairs, $\pazocal{M}$—black solid line—increases to reach the maximal multiplicity at an altitude of, roughly, one stellar radius ($R_*$) from the surface. The acceleration of the charges, expressed in units of $\gamma_p$—black dashed line—, shows a radial profile compatible with pair multiplicity. The wavelength of the photons resulting from the resonance condition—$\omega_p\sim m_a$—for this star model is shown with a blue dash-dot line. As a reference, the resonant mass of axions converted into photons within the GJ model framework—$\omega_p\sim m_a^{\mathrm{GJ}}$—is shown with a thin dashed blue line. The shaded region in gray corresponds to the polar cap and should not be considered within the framework of this study. }
\label{fig_R1}
\end{figure}
We rely on Refs. \cite{Timokhin:2015dua, Timokhin:2018vdn} to draw a minimal young pulsar model as shown in Fig. \ref{fig_R1}: (i) the multiplicity is close to GJ in regions where the induced electric fields that accelerate pairs are unscreened by the plasma, i.e. $\pazocal{M}\sim1$, for $r/R_*\lesssim 1/100$; $\pazocal{M}$ reaches the maximal multiplicity between $0.001\lesssim r/R_* \lesssim 0.5$; $\pazocal{M}$ maintains its maximal value for $r/R_*\gtrsim 0.5$, (ii) the plasma Lorentz factor fits $\gamma_p \sim 1$ for $r/R_*\lesssim 0.03$; $\gamma_p \sim 10^3$ for $0.03 \lesssim r/R_*\lesssim 0.08$; $\gamma_p \sim 10^4$ for $0.08\lesssim r/R_*\lesssim 0.1$; and $\gamma_p \sim 10^5$ at $r/R_*\sim 1$. 

Although in this section the propagation is constrained to $\theta\approx\pi/2$ for simplicity, which leads naturally to the parallel propagation of O-modes while L-O modes gradually vanish, once we confine our interest to the vicinity of open field lines this is a good approximation for most pulsars as they should be, roughly speaking, aligned rotators. Indeed, the uncertainty introduced by considering most of input parameters—$\omega_p$, $\pazocal{M}$, $\gamma_p$, etc.—to vary smoothly over the conversion region, which could be realistic and has been stated before by the literature \cite{PhysRevLett.121.241102}, is also much smaller than the inherent uncertainty in the input astrophysical parameters—the velocity of baby pairs, differential acceleration and turbulence, scattering, etc.—together with QED corrections and, hence, does not detract from the interest of our study. Moreover, \ref{Ap3} sketches a new 3D model that could potentially address some of these points, thereby fine-tuning the calculations.

The polar cap radius is $r_{\mathrm{pc}}\sim R\sqrt{R/r_\mathrm{L}}$, with $r_\mathrm{L}\sim c/\Omega_*$ being the light cylinder radius. This corresponds to $r_{\mathrm{pc}}\sim R_*/25$ for a Geminga-like (PSR J0633+1746) pulsar—$P\simeq24$ ms—, or $r_{\mathrm{pc}}\sim R_*/40$ for Crab (PSR B0531+21)—$P\simeq33$ ms. Throughout this section, we excise the polar cap of the NS to focus on higher altitudes from the surface. As will be shown, this would potentially broaden the spectrum of interest, probing a wider variety of heavier axions with respect to previous expectations, which focused on examining magnetospheric regions where the GJ model may be a reasonable approximation. To explore the wavelength of the axion-induced photons at high altitude, we have calculated the resonance frequency for the simplified pulsar model shown in Fig. \ref{fig_R1}. In Fig. \ref{fig_R1}, we show the wavelength of the photons resulting from the resonant conversion of DM axions varies from mm to cm for $0.01<r/R_*<2$ in a blue dash-dot line. As a reference, the resonant mass of axions converted into photons in the GJ framework is shown with a thin blue line, which is significantly lower, always within the centimeter range of microwaves. This seems to support the hypothesis that axions of higher masses may resonate in certain magnetospheric splits, which would translate into an axion-induced spectral feature at frequencies of up to roughly 100 GHz.

Now, we can address the question of whether such a high-frequency signal is detectable by realistic experiments. Telescopes integrate over the angle subtended by a surface in the sky as $\mathrm{S}_\nu \approx \int I_\nu(\theta, \phi) \, d\Omega$, where $I_\nu$ is the intensity per unit frequency and $\theta$, $\phi$ are the polar and azimuth angle, respectively. In turn, for smaller conversion zones in comparison to the star size, the solid angle subtended by the conversion regions in the vicinity of open field lines—i.e., above the polar cap, etc.—is smaller, and hence the spectral line originating from axion-photon resonance dilutes with the time average of the so-called dilution factor as $\mathrm{S}_{\nu}\propto \Omega_c / \Omega$, with $\Omega_c$ being the solid angle associated with the conversion zone. The signal suppression factor can cause geometric dilution by several orders of magnitude, as the projection of the open field lines is smaller than the region defined by the closed field lines or the radio beam. Magnetospheric splits with high $\pazocal{M}/\gamma_p$ are associated with pulsed emission near the magnetic axis—see Fig. \ref{fig_R2}. During observations, data is accumulated from regions responsible for either pulsed or steady emission, depending on the pulsar's rotational angle, and can therefore be separated using Fourier transform. 

Although the literature suggests that the magnetosphere contains `large'  volumes where pair creation cannot effectively screen the electric field \cite{Timokhin:2015dua, Timokhin:2018vdn}, determining $\Omega_c$, the solid angle subtended by the magnetospheric region compatible with the axion resonance for masses closer to the meV, remains an ongoing area of research in astrophysics. On the other hand, as inferred from inputting values from \ref{A1}, Table \ref{table1} into Eq. \ref{Eq.7}, cataloged isolated pulsars are too distant to generate an axion-induced signal detectable by ground-based observatories, even for $\Omega_c/\Omega\sim1$. Once signals from known isolated sources appear too weak, one possibility is to consider pulsar populations \cite{PhysRevLett.121.241102, Safdi:2018oeu, PhysRevLett.125.171301, Foster:2022fxn}. However, a naive estimate using Eq. \ref{Eq.9} suggests that the QCD axion could be detectable in millimeter-wave observations of a population of $n \gtrsim 10^5$ young to middle-aged pulsars in the galaxy's inner regions \cite{Foster:2022fxn}, producing a flux density on Earth of $\sim$1/2 mJy, enhanced by the DM overdensity in the GC \cite{McMillan_2016}, for a dilution factor threshold of $\Omega_c/\Omega > 10^{-3}$. Even if these conditions arise in nature, it should be noted that methods for separating pulsed and quiescent emission in isolated pulsars are well-established—e.g., \cite{Torne:2015rha}—, but applying them to pulsar populations may quickly become unmanageable. Therefore, while it seems plausible that heavier axion dark matter could resonate in pulsars and their populations, it remains unclear how long the associated signal can be detected from Earth.

\section{Magnetars} \label{Ap2}
It is straightforward to plug the values from \ref{A1}, Table \ref{table1}, into Eq. \ref{Eq.7} and observe that most known magnetars will produce an axion spectral feature that is hardly detectable. However, as shown in Ref. \cite{McMillan_2016}, the central region of the Galaxy is a bulk where an axion DM density of up to $\rho_{\infty}\sim10^8$ GeV cm$^{-3}$—i.e., up to $\sim$$10^9$ times denser than our near halo—is tenable. Since the signal induced by ambient axions falling into the star and being converted into photons scales with the DM density at the position of the object—S$_\nu \propto\rho_{\infty}$—, the magnetar SGR 1745--2900, with its strong magnetic field of about 1.6$\times 10^{14}$ G at surface \cite{Mori:2013yda}, is thus expected to be an efficient axion-to-photon converter in the GC. These aspects should largely compensate for the number of observational hurdles—$\sim$8 kpc distance, SGR 1745--2900 is about 0.1 pc apart from SgrA*. Currently, SGR 1745--2900 shows a weak radio continuum; and presents time variability, likely approaching a new quiescent phase \cite{10.1093/mnras/stx1439}. As a benchmark, Ref. \cite{Darling:2020uyo} reports a continuum of SGR J1745--2900 of some 20--5 mJy in L to Ka bands—about 1--40 GHz—in year 2014; while Ref. \cite{10.1093/mnras/stx1439} found a SGR 1745--2900 continuum peak intensity of $\sim$4 mJy beam$^{-1}$ at 226 GHz in 2017. SGR 1745--2900 and SgrA* are about 2.4 arcsec apart. SgrA* shows an integrated flux density of some 2 Jy in centimeter band,  and a peak intensity of about 3 Jy beam$^{-1}$ in millimeter band \cite{10.1093/mnras/stx1439}. Despite this observational challenge, Refs. \cite{Darling:2020uyo, 10.1093/mnras/stx1439} shown that it is possible to resolve out the Galactic Center, separating SgrA* and SGR 1745--2900 continuum in cm to mm band.
\begin{figure}[h]\centering
\includegraphics[width=.5\textwidth]{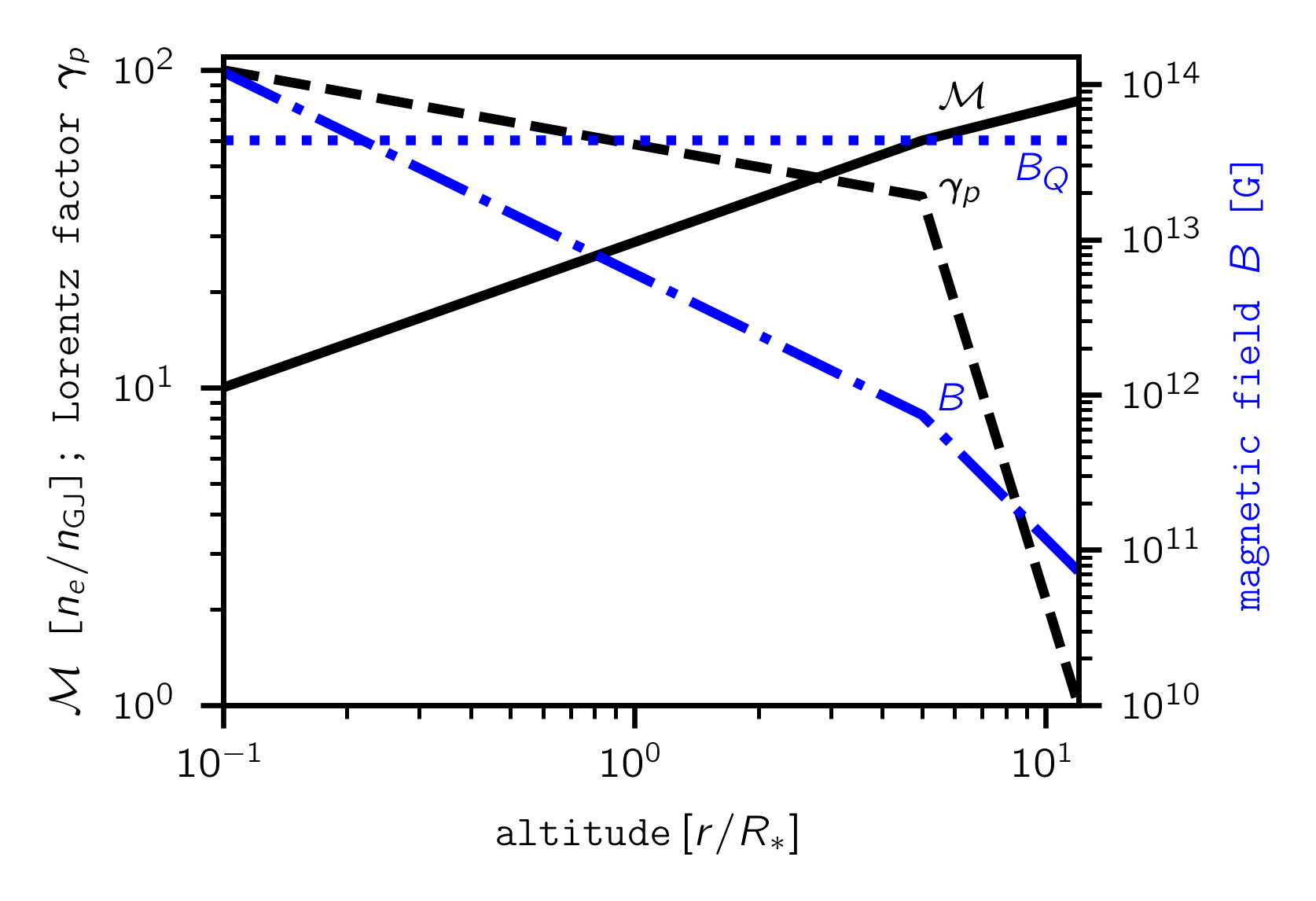}
\caption{Toy model of magnetar as inspired by the results from Beloborodov \cite{Beloborodov_2012, Beloborodov_2013} for $0.1<r/R_*<12$. The multiplicity of pairs, $\pazocal{M}$, is $\pazocal{O}(50)$ within the significant part of the magnetospheric splitting. The acceleration of the charges is measured in terms of $\gamma_p$. The conversion of photons into electrons is mitigated at $B\sim10^{13}$ G but the steady relativistic outflow maintains $\pazocal{M}$ roughly constant at higher altitudes. The dipole magnetic field—blue dash-dot line—is modeled in the weakly twisted limit. The critical field, $B_Q$, is represented by a thin dashed blue horizontal line for reference.}
\label{fig_R3}
\end{figure}

\begin{figure}[h]\centering
\includegraphics[width=.5\textwidth]{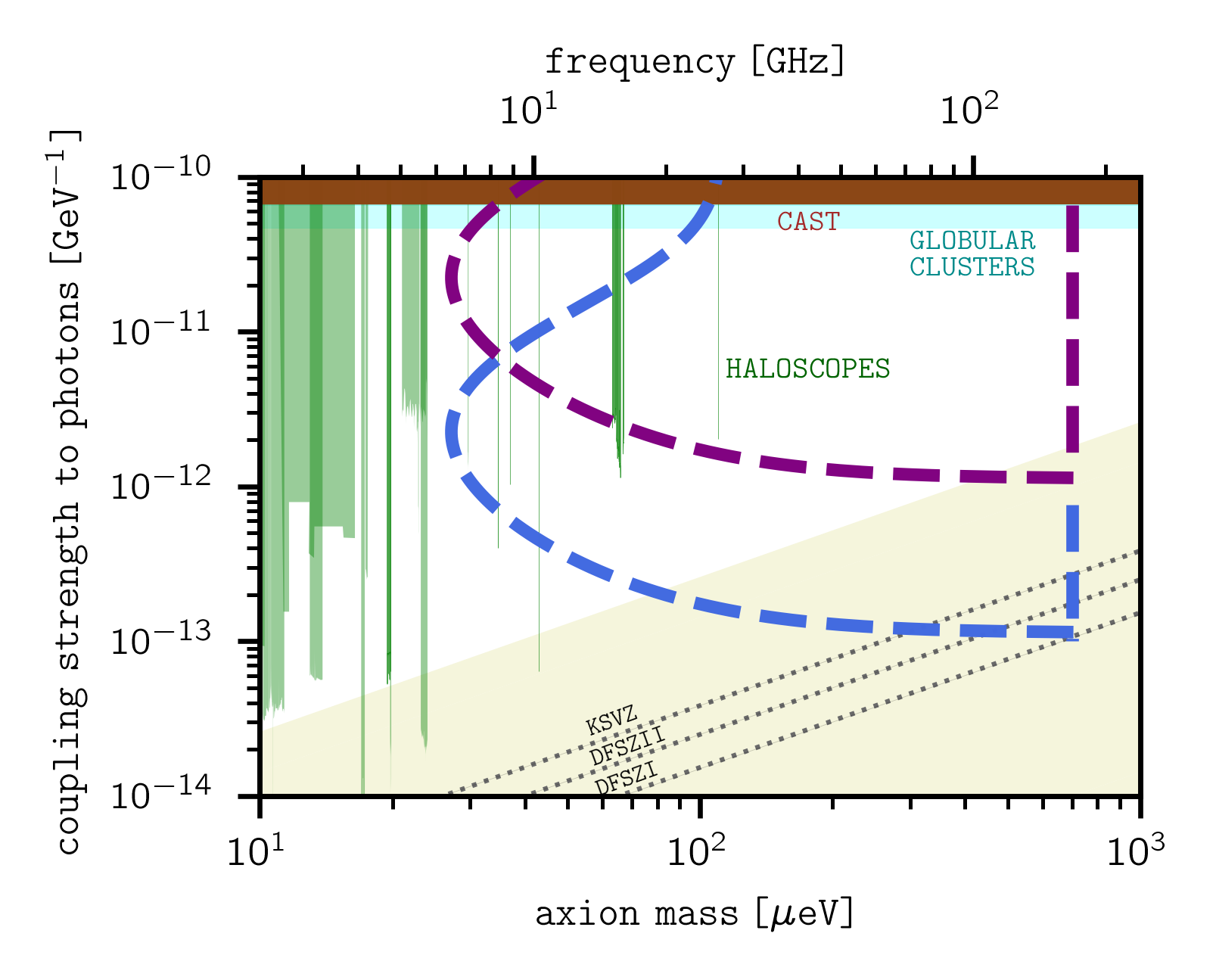}
\caption{Accessible axion-photon coupling strength in observations of SGR 1745–2900 employing the toy magnetar model shown in Fig. \ref{fig_R3} for a threshold flux density on Earth of $\sim$1/2 mJy. Other star parameters are taken from Table \ref{table1}. The dilution factor is $\Omega_c/\Omega \sim 1$.  The ripple of the projections that allows two different sensitivities for some axion masses is a consequence of the crossing of the radial profiles of pair density and plasma Lorentz factor with slopes of opposite signs that enable distinct slots with a same $\pazocal{M}$/$\gamma_p$ ratio. In \cite{Darling:2020plz}, $6\times10^4\lesssim \rho_{\infty}\mathrm{/GeV/cm^{-3}}\lesssim6\times10^8$ is considered for SGR 1745--2900, based on \cite{McMillan_2016, PhysRevLett.121.241102}, which we adopt in this graph. Thus, two ambient DM densities of $\rho_{\infty}\approx 6\times10^6$ $\mathrm{GeVcm^{-3}}$—purple lines—and $6\times10^8$ $\mathrm{GeVcm^{-3}}$—blue—are considered. Simulations are projected onto current exclusion sectors. The QCD window is defined by benchmark axion models \cite{PhysRevLett.43.103, Shifman1980CanCE, DINE1981199, osti_7063072, DiLuzio:2016sbl}. The \textit{phenomenologically preferred} axion window is shaded in beige \cite{DiLuzio:2016sbl}. The white sector is compatible with axion-like particles \cite{doi:10.1146/annurev-nucl-120720-031147, doi:10.1126/sciadv.abj3618}. A detection in the reachable sector could be corroborated by direct methods using new-generation haloscopes—e.g., see  \cite{DeMiguel:2023nmz} and references therein for a recent overview.}
\label{fig_1}
\end{figure}

We rely on Refs. \cite{Beloborodov_2012, Beloborodov_2013} to build up a minimal magnetar model as shown in Fig. \ref{fig_R3}:
the rate of resonant scattering by relativistic charges that boosts multiplicity increases as particles ascend from $B\gg B_Q$ to $B \lesssim B_Q$. As a result, the pair multiplicity of the outflow rises from $\pazocal{M} \sim 1$ to $\pazocal{M} \sim 100$. Although charges creation is mitigated at $B \sim 10^{13}$ G, the steady relativistic outflow maintains $\pazocal{M}$ roughly constant at higher altitudes along the field lines. For $\pazocal{M}\sim50$, the plasma Lorentz factor is $\gamma_p\sim100$ for $r/R_*\lesssim2$; $\gamma_p\sim10$ for $2\lesssim r/R_*\lesssim 10$; and $\gamma_p\sim1$ for $r/R_*\gtrsim10$. According to \cite{Beloborodov_2012, Beloborodov_2013}, the relevance of the angle perceived by the observer during an observation may fade in comparison to the case of pulsars discussed in Sec. \ref{III} since $\gamma_p$ evolves from $\gamma_p\sim1$ to $\gamma_p \gg 1$ independently of the angle along a radial profile for a given $\pazocal{M}$. On the other hand, it has been suggested that the magnetic field may fall with altitude as $B\propto r^{-(2+p)}$ with $p<1$ in magnetars \cite{Thompson:2001ig}. The model developed throughout Sec. \ref{II} does not address corrections for the twisted fields of magnetars. However, following the example of \cite{Beloborodov_2013}, simulations with SGR 1745--2900 are here performed in the limit of a weakly twisted field, i.e. $p\approx1$. Canonical dimensions for radius and mass are adopted for simulations of SGR 1745--2900 in Fig. \ref{fig_1}. In Fig. \ref{fig_1}, the broadening of the band of interest from the microwave band to the millimeter range is noticeable. Interistingly, axion models inspired by QCD, with masses close to the meV, emerge as detectable from Earth under an optimistic scenario.

\section{Summary and conclusions}\label{IV}
In this work, a simple model is built-up to forecast the intensity of the narrow-band spectral feature originating from the conversion of ambient DM axions in splits of the highly magnetized relativistic plasma of NS magnetospheres where the electromagnetic cascade plays an important role. 

This pathfinding model accommodates, for the first time, in a simultaneous manner (i) the pair multiplicity factor, and (ii) the Lorentz factor of the plasma; drawing interesting conclusions: there may be regions of the magnetosphere with diverse dynamics and occupancy. Lighter and heavier axions would penetrate and resonate in distinct zones, producing an axion-induced emission line somewhere between the radio domain and the millimeter-wave range. 

As noted in Sec. \ref{III}, it remains uncertain whether any high-frequency spectral features from pulsars, or their populations, can be observed from Earth. This is due to a combination of factors such as distance, signal dilution caused by the smaller size of the regions resonating at high frequencies, and observational challenges. In the case of magnetars, QED could govern the magnetosphere dynamics, causing distortions of the radial functions of the input parameters of the model. In spite of that, the simulations performed in the limit of a weakly twisted field for the magnetar SGR 1745--2900 in Sec. \ref{Ap2} are a promising sign of plausible millimeter-wave signals that could emerge from such intriguing objects. In future work, we aim to present refined simulations after further developing the model shown in \ref{Ap3}.

To recapitulate and conclude, the main result of this article reads as follows: heavier axion DM could mix with photons in the magnetosphere of NSs up to, say, a meV in mass—frequencies of a few hundred GHz—, in contrast with previous expectations, since the relativistic plasma density is enhanced by charges creation, and hence the apparent mass of the resonant photon, which scales with the ratio between the pair multiplicity and the Lorentz factor in the conversion zone roughly as $\omega_p\propto\sqrt{\pazocal{M}_c/\gamma_c}$. Consequently, the range of wavelengths for which axion field resonance is tenable in neutron star magnetospheres can span from the microwave domain—for $\pazocal{M}_c/\gamma_c\sim1$—to the millimeter-wave range—if $\pazocal{M}_c/\gamma_c\gg1$. This would allow a broader test to the DM axion through multi-frequency observations than previously expected, since observational work to date looks for the axionic signal below some 40 GHz. In particular, the case of SGR1745--2900 is promising. In this regard, there are several atmospheric windows at frequencies of a few dozen/hundred GHz. We have confirmed through simulations using the observing tools \cite{10.1117/12.789108, heywood2011almaobservationsupporttool} that the current generation of ground-based observatories is sensitive to axion-induced line intensities of a fraction of mJy in the millimeter band,\footnote{Among others, the Atacama Large Millimeter/submillimeter Array (ALMA) and The Submillimeter Array (SMA).} which gives us the means to ensure that searches for axion-caused spectral features at higher frequencies receive well-deserved attention.

\section*{Acknowledgements}
The project that gave rise to these results received the support of a fellowship from “la Caixa” Foundation (ID 100010434). The fellowship code is LCF/BQ/PI24/12040023”. J.D.M. acknowledges support from the Spanish Ministry of Science, Innovation and Universities and the Agency (EUR2024-153552 financed by MICIU/AEI/10.13039/501100011033). This work was supported by RIKEN’s program
for Special Postdoctoral Researchers (SPDR)—Project
Code: 202101061013. The comments of Z. Wadiasingh and A. Millar are gratefully acknowledged. Thanks F. Poidevin, E. Hatziminaoglou, and N. Rea for discussions.

\appendix

\section{Casuistic}\label{A1}
This appendix explores the diversity of Galactic sources to which the model in Sec. \ref{II} may be transferable. Canonical dimensions of a NS are $M_*\sim10$ km and $R_*\sim1\,\mathrm{M_{\odot}}$, and small deviations would not significantly alter the conclusions drawn in this article; while the rotation period ranges from tens of milliseconds to a dozen seconds, approximately. Plasma anisotropies and $\pazocal{M}$ factor insertion plus relativistic corrections, measurable through the Lorentz factor characteristic of the plasma, $\omega_p$, are barely known. In this respect, three NS classes amply confine our attention: \\(i) observations are available throughout almost the entire electromagnetic spectrum for young, rotation powered pulsars, e.g., Crab (PSR B0531+21) and middle-aged, e.g. Geminga (PSR J0633+1746) or Vela (PSR B0833–45); the flux density on Earth from the latter ranging from a Jy at radio frequencies to a few $\upmu$Jy at millimeter wave, as reference—cf. \cite{Mignani:2017api}. They are also the objects that have been most extensively discussed in the literature, and there are studies that determine some of their basic characteristics, as well as simulations that establish a density profile compatible with data, whose intersection suggests that the multiplicity should not exceed $10^4\lesssim\pazocal{M}\lesssim\mathrm{few}\times10^5$ \cite{Timokhin:2015dua, Timokhin:2018vdn,Lyutikov:2007xw, 2007ApJ...658.1177D, Olmi:2016avl, 10.1111/j.1365-2966.2010.17449.x}. In terms of simulations, it is emphasized that the model comes under stress in the modeling of millisecond pulsars; \\(ii) so-called `anomalous x-ray pulsars' and `soft gamma repeaters' (SGRs) are thought to be highly magnetic NSs, referred to as magnetar \cite{Duncan1992FormationOV}. Submillimeter wave astronomy of magnetar XTE J1810--197 reveals a quiescient flux density of several mJy at few hundred GHz and sets 3$\sigma$ upper limits of several mJy at about six hundred GHz, while at microwave frequencies the flux density is of the order of a few mJy \cite{Torne:2022xfo}; which is in line, in broad terms, with the multi-frequency astronomy of SGR 1745--2900 \cite{Torne:2015rha, Torne:2016zid, 10.1093/mnras/stx1439}. The x-ray spectrum of magnetars cannot be explained in terms of rotational energy losses \cite{Thompson:2001ig}. Measurement of the spin period and its time derivative suggest that these objects are endowed with super strong polar fields of about $B_p =\mathrm{several}\times 10^{19}(P\dot{P})^{1/2}\sim 10^{14-15}$ G, as calculated in the limit for a simple magnetic dipole braking in vacuo. Fields in excess of the critical field, $B_Q = m^2_e/e\approx4.4\times10^{13}$ G, above which photon splitting starts affecting cascade multiplicity, would mitigate pair creation \cite{Baring:2000cr}. The density profile in the emission region of some of those objects has been analyzed by confronting simulations with x-ray observational data in \cite{Rea:2008zs}, concluding $\pazocal{M}\lesssim10^3$, which agrees with \cite{ Beloborodov_2013, Beloborodov_2012}. We emphasize that the radial profile evolved from a GJ-like model of Eq. \ref{Eq.2} could be a too simplistic approximation in the case of magnetars with pronouncedly twisted fields, since they could scale as $B\propto r^{-(2+p)}$ with $p<1$ instead of $B\propto r^{-3}$ as regular pulsars \cite{Thompson:2001ig}; \\(iii) binary systems composed of a NS and a black hole (NS--BH) and double pulsars (NS--NS) are also incorporated to Table \ref{table1}, in which we use a Navarro--Frenk--White (NFW) density profile—$\rho (r)={ {\rho _{\rm {halo}}}/[{3A_{\rm {NFW}}\,x(1/C+x)^{2}}]}$; with $\rho _{\rm {halo}}= M/(4\pi/3 r_{\rm {vir}}^{3})$, $M=4 \pi \rho _{\odot} r_s^3 A_{\rm {NFW}}$, $r_s\sim 20$ kpc, $r_{{\mathrm  {vir}}}=Cr_{s}$, $A_{\rm {NFW}}=\ln(1+C)-C/(1+C)$, $C\sim10$, $x=r/r_{\rm {vir}}$—for the estimation of $\rho_{\infty}$ in the case of isolated sources distant from the GC \cite{Navarro:1995iw}. Differently, in the case of the magnetar SGR 1745--2900, in orbit around the central BH in the Milky Way (MW), Sagittarius A* (SgrA*), huge overdensities of DM with respect to the typical occupation number of particles, which would increase the axion-induced emission accordingly, are expected. In \cite{Darling:2020plz}, $6\times10^4\lesssim \rho_{\infty}\mathrm{/GeV/cm^{-3}}\lesssim6\times10^8$ is considered for SGR 1745--2900, based on \cite{McMillan_2016, PhysRevLett.121.241102}, which we adopt in this manuscript. In this line, \cite{Rollin_2015}  suggests a capture of DM particles of about four orders of magnitude with respect to the ambient occupation, due to gravitational focusing, in the vicinity of a supermassive BH. On the other hand, in the case of double pulsars, such as PSR J1906+0746 or PSR J0737–3039, the enhancing factor is weaker, and depends on the orbital period, resulting in the captured DM density being a few--several times the environmental density at the position of the system in the Galaxy due, mainly, to slingshot \cite{PhysRevLett.109.061301}. Other relevant parameters are also determined by observation of the eclipse of binaries that are more difficult to infer for isolated NSs \cite{Lyutikov:2005zw, Lyne}. Interestingly, simulations strongly suggest that the plasma density exceeds the GJ density by a factor of up to $10^{4-5}$ in PSR J0737--3039 \cite{Lyutikov_2005}.

\begin{table*}
    \begin{minipage}{\textwidth}
%    \centering
%\begin{table*}
\centering
\caption{Simulation parameters and results. Pulsars are grouped in the first half; magnetars after the horizontal line; binaries are marked with a double dagger.$^{(\ddagger)}$  $M_*\sim1.4$ $\mathrm{M_{\odot}}; R_*\sim10$ km; $\theta_m\sim10^\circ$; $v_0\sim200$ km s$^{-1}$; $\theta\sim90^\circ$ are adopted in the absence of data. The local DM density is set $\rho_{\odot}\sim0.4$ GeV cm$^{-3}$.}
\label{table1}
\resizebox{\textwidth}{!}{%
\begin{tabular}{cccccccccccc}
\toprule
\multirow{2}{*}{Source}    & \begin{small} $M_*$ \end{small}                  & \begin{small}$R_*$  \end{small}               &  \begin{small}   $B_0$ \footnote{$B_p \sim B_0$ is adopted without adding uncertainty, as the field at surface, $B_0$, and the polar cap field, $B_p$, are of the same order \cite{CotiZelati:2017rgc,Gogus:2020coh}.}          \end{small}            &\begin{small} $P$     \end{small}     & \begin{small} $\theta$  \end{small}           & \begin{small} $\theta_m$  \end{small}               &\begin{small}  $v_0$      \end{small}  &\begin{small}  $d$      \end{small} &\begin{small}  $\rho_{\infty}$      \end{small}               & \multirow{2}{*}{$\pazocal{M}$ } & \multirow{2}{*}{Ref.}\\
& \multicolumn{1}{c}{\begin{footnotesize}/$\mathrm{M_{\odot}}$\end{footnotesize}} & \multicolumn{1}{c}{\begin{footnotesize}/km\end{footnotesize}} & \multicolumn{1}{c}{\begin{scriptsize}/G/$10^{14}$\end{scriptsize}} & \multicolumn{1}{c}{\begin{footnotesize}/s\end{footnotesize}} & \multicolumn{1}{c}{\begin{footnotesize}/deg\end{footnotesize}} & \multicolumn{1}{c}{\begin{footnotesize}/deg\end{footnotesize}} & \multicolumn{1}{c}{\begin{footnotesize}/km$/$s$^{-1}$\end{footnotesize}} &  \multicolumn{1}{c}{\begin{footnotesize}/kpc\end{footnotesize}} &  \multicolumn{1}{c}{\begin{footnotesize}/$\!\rho_{\odot}$\end{footnotesize}} & & \\
\midrule
PSR J0633+1746 (Geminga)       & 1.4                                        & 10            &   .033            &  .237   & 90    & 10                            & 200   &.25  & 1.1& $\lesssim$$10^5$  &  \cite{ Timokhin:2015dua, Timokhin:2018vdn, Lyutikov:2007xw, 2007ApJ...658.1177D, Olmi:2016avl, 10.1111/j.1365-2966.2010.17449.x}  \\
PSR B0833--45 (Vela)                & 1.4                                         & 10          & .068            & .089  & 90  & 10                            & 200   &.28  & 1.1 & $\lesssim$$10^5  $   &  \cite{ Timokhin:2015dua, Timokhin:2018vdn, Lyutikov:2007xw, 2007ApJ...658.1177D, Olmi:2016avl, 10.1111/j.1365-2966.2010.17449.x}     \\
PSR B2334+61                  & 1.6                                       & 13          & .1            & .495  & 90  & 10                            & 200  &3.1  & .8 & $\lesssim$$10^5 $    &   \cite{ Timokhin:2015dua, Timokhin:2018vdn, Lyutikov:2007xw, 2007ApJ...658.1177D, Olmi:2016avl, 10.1111/j.1365-2966.2010.17449.x, McGowan:2005kt}   \\
PSR B0656+14                  & 1.4                                         & 10          & .093            & .385  & 90  & 10                            & 200  &.28  & 1.0 & $\lesssim$$10^5  $     &  \cite{ Timokhin:2015dua, Timokhin:2018vdn, Lyutikov:2007xw, 2007ApJ...658.1177D, Olmi:2016avl, 10.1111/j.1365-2966.2010.17449.x}   \\
PSR J0737--3039 B$^{(\ddagger)}$                 & 1.25                                         & 10          & .02            & 2.77  & 90.5  & 15                            & 323\footnote{System velocity plus orbital velocity of B should be computed. This is left for future work with the modeling of the line broadening. }  &1.15  & 3--4\footnote{Cf. \cite{PhysRevLett.109.061301} for the analysis of DM capture in PSR J1906+0746; transferable to PSR J0737--3039 of known $\pazocal{M}$ leading order  \cite{Lyutikov_2005}.} & $\lesssim$$10^5  $  &   \cite{ Timokhin:2015dua, Timokhin:2018vdn, Lyutikov:2007xw, 2007ApJ...658.1177D, Olmi:2016avl, 10.1111/j.1365-2966.2010.17449.x,Lyutikov_2005} \\
\midrule
1RXS J170849.0--4009       & 1.4                                        & 10          & 9.3           & 11.01   & 90 & 10                           & 200 &3.8   &  2.6 & $\lesssim$$10^3$   &  \cite{ Rea:2008zs}                 \\
4U 0142+614              & 1.4                                        & 10          & 2.7           & 8.69  & 90  & 10                           & 200  &3.6   & .7 & $\lesssim$$10^3  $ &  \cite{  Rea:2008zs}\\
1E 1841–045              & 1.4                                        & 10          & 13.8          & 11.79  & 90  & 10                           & 200  &8.5  &  3.2 & $\lesssim$$10^3  $ & \cite{  Rea:2008zs} \\
1E 2259+586              & 1.4                                        & 10          & 1.2           & 6.98  & 90  & 10                           & 200  &3.2   & .8 & $\lesssim$$10^3$  &  \cite{  Rea:2008zs} \\
1E 1048–5937             & 1.4                                        & 10          & 7.6           & 6.46  & 90  & 10                           & 200  &9  & .8  & $\lesssim$$10^3 $  &  \cite{  Rea:2008zs}  \\
XTE J1810–197           & 1.4                                        & 10          & 2.6           & 5.54  & 90  & 10                           & 200   &3.5   & 2.4 & $\lesssim$$10^3$  &\cite{  Rea:2008zs}    \\
1E 1547.0–5408           & 1.4                                        & 10          & 6.4           & 2.91  & 90  & 10                           & 200  &4.5   & 2.3 & $\lesssim$$10^3 $  & \cite{  Rea:2008zs}  \\
SGR 1806–20               & 1.4                                        & 11.57       & 40            & 7.55  & 90  & 10                           & 200   &8.7    & 10.2 & $\lesssim$$10^3  $ & \cite{  Rea:2008zs, Colaiuda:2010pc} \\
SGR 1900+14              & 1.4                                        & 10.91       & 42            & 5.20  & 90  & 10                           & 200   &12.5  & 1.0& $\lesssim$$10^3 $   &   \cite{ Rea:2008zs, Colaiuda:2010pc}  \\
SGR 1745--2900$^{(\ddagger)}$ & 1.4                                          & 10          & 1.6           & 3.76  & 90  & 10                            & 236  &8.3  &  $10^5$  & $\lesssim$$10^3 $\footnote{I hypothesize that this object presents pair production similar to that of other magnetars in \cite{Rea:2008zs}, with upper-limit $10^2\lesssim\pazocal{M}\lesssim10^3$ \cite{ Beloborodov_2013}.} &  \cite{ Beloborodov_2013, Mori:2013yda, Bower:2014tea}  \\ 
 &                                           &           &            &   &   &                             &  &  &  $10^9$\footnote{Two simulations that can be thought as lower/upper limit obtained from density profiles in \cite{PhysRevLett.121.241102, Darling:2020plz, Rollin_2015}.}  &  & \cite{PhysRevLett.121.241102, Darling:2020plz, Rollin_2015} \\
\bottomrule
\end{tabular}%
}
\end{minipage}
\end{table*}

In contrast, the following sources are excluded from Table \ref{table1}:
\\(iv) we would not consider for simulations `central compact objects,' likely NSs whose properties differ from pulsars and magnetars \cite{Gotthelf:2007mm}, as the characteristic field at surface is significantly weaker, of the order of $B\lesssim3\times10^{11}$ G, and the conversion rate would be reduced accordingly; \\(v) there is a region between the fiduciary properties of rotation-powered pulsars and magnetars, at which a NS exhibits magnetic field around the critical field and rotation periods of the order of a second, i.e., `high magnetic field pulsars,' which the literature has neglected at the time of performing simulations to determine the pair multiplicity factor, perhaps with the exception of \cite{Timokhin:2015dua, Timokhin:2018vdn}, at which fields up to $\lesssim 3\times 10^{13}$ G and periods up to a fraction of second are considered. Therefore, we exclude these objects even though it could be conjectured that they present maximal multiplicities within the range $10^3\lesssim\pazocal{M}\lesssim10^5$ by simple extrapolation; \\(vi) so-called `x-ray dim isolated NSs' (XDINs) are rare objects. Their longer spin period, ranging from a few seconds to a dozen, are an indicator of high magnetic fields, of the order of $10^{13}$ G, required to decelerate the rotation in a time frame compatible with star age \cite{Haberl:2006xe}. The preeminent consequence of any suppression of pair creation in pulsars is that the emission of radio waves should be strongly inhibited \cite{Baring:2000cr}. That could explain the lack of detected radio emission from XDINs, with upper limit of $\sim$5 mJy at submillimeter wave \cite{10.1111/j.1365-2966.2010.16557.x}; or they could simply be misaligned long-period nearby radio pulsars with high magnetic fields beaming away from the Earth, while alternative models to GJ have been proposed in order to solve the discrepancies \cite{Keane:2008jj, Kondratiev_2009}. In spite of the obvious observational advantage that the seven known XDINs represent, specially if we note their proximity to Earth, of a few hundred pc distance, I cautiously leave XDINs outside the scope of this analysis in the absence of a well-defined framework.

In Table \ref{table1} a number of active sources for which the relevant parameters can be determined, or approximated, are collected, for which we rely, mainly, in \cite{CotiZelati:2017rgc}. We emphasize that the reduced catalogue in Table \ref{table1} is susceptible to incompleteness, and biased, nevertheless providing one with a tool that enables the qualitative study within the scope of this article to be continued towards observational studies. In this concern, from the simple exercise of inserting real star parameters into Eq. \ref{Eq.7}, it becomes rapidly clear that most sources would produce a faint axion-induced signal hardly detectable by telescopes,  orders of magnitude weaker than a $\upmu$Jy. Two potential exceptions are discussed in the main letter.

\section{A step towards a compelling 3D model}\label{Ap3}
The model in \cite{ Millar:2021gzs} propagates oblique—L-O—modes in the strongly magnetized relativistic plasma of NSs. This model can be modified to accommodate the relativistic plasma frequency, $\omega_p$, and the multiplicity of charges, $\pazocal{M}$, in order to perform 3D simulations with a greater variety of propagation directions. This would also enable the fine-tuned calculation of the photon flux induced by DM axions in real NSs that are imperfectly aligned. In this concern, the `flux transfer' in a WKB approximation, that can be understood as a the axion-photon conversion rate, is
$R \simeq g^2_{a \gamma} B^2(r) / (2 k |\omega'_p|) \times \pi m^5_a \mathrm{sin}^2\theta/(k^2+m^2_a \mathrm{sin}^2\theta)^2$ \citep{ Millar:2021gzs}. Here, two terms must be derived before evaluating $R$. First, the dispersion relation for the L-O modes that determines $k$ can be directly addressed with Eq. \ref{Eq.10} \cite{PhysRevE.57.3399}. Second, the gradient of the relativistic plasma frequency, $|\omega'_p|$, can be determined through the operator \cite{ Millar:2021gzs}
\begin{equation}
\begin{aligned}
(...)'=\frac{\partial(...)}{\partial r} = \frac{\partial(...)}{\partial z} + \frac{\omega^2_p \xi}{\omega^2 \mathrm{tan}\theta}  \frac{\partial(...)}{\partial y}
\;,
\label{Eq.A2}
\end{aligned}
\end{equation}
where $\xi=\mathrm{sin^2\theta}/(1-\omega^2_p\mathrm{cos^2\theta/\omega^2})$, with 
\begin{equation}
\begin{aligned}
\omega^2_p=\frac{m^2_a \omega^2}{m^2_a(z/r)^2+\omega^2(y/r)^2}
\,,
\label{Eq.A3}
\end{aligned}
\end{equation}

where $\hat{r}=\mathrm{sin}\theta \hat{y}+\mathrm{cos}\theta \hat{z}$. Note, if axions are poorly relativistic, $\omega\sim m_a$ is accurate, and hence Eq. \ref{Eq.A3} simplifies to $\omega_p\sim m_a$. The sub-index `c' referring to the conversion region is omitted throughout this appendix for simplicity, while the sign and angle convention as defined in \cite{ Millar:2021gzs} is used.

In this sketch, the plasma Lorentz factor plays a role through the dispersion relation in Eq. \ref{Eq.10} as, locally, $\langle \gamma_p ^{-3} \rangle \sim \gamma_p^{-1}$, while the role of the multiplicity factor can be brought to surface, for example, by clearing out $B(r)$ in Eq. \ref{Eq.2} with Eq. \ref{Eq.3} in order to insert it into $R$, obtaining 
\begin{equation}
\begin{aligned}
R\simeq \frac{g^2_{a \gamma}} {2 k |\omega'_p|} \left(\frac{e}{2\Omega_*}\frac{m_e}{4\pi\alpha}\frac{\gamma_p}{\pazocal{M}}\omega^2_p \right)^2 \frac{\pi m^5_a \mathrm{sin}^2\theta}{(k^2+m^2_a \mathrm{sin}^2\theta)^2}
\;.
\label{Eq.A4}
\end{aligned}
\end{equation}
In the case of adopting a dipole field approximation with  $B(r)=1/2B_0(R_*/r)^3f_1(t)$ as in Sec. \ref{II}, from Eq. \ref{Eq.3} we find $\omega'_p=\sqrt{4\pi\alpha\Omega_*B_0 R^3_*\pazocal{M}/(m_e e\gamma_p)}\,\partial( r^{-3/2}f^{1/2}_1(t))/\partial r$ for an arbitrary polar angle; while the conversion altitude at which $\omega_p\sim m_a$ could be obtained from Eq. \ref{Eq.4}. Of course, the resulting model would not accommodate distorted non-radial field distributions, so further corrections may be necessary to perform simulations with magnetars—for instance, considering a magnetic field scaling more smoothly with altitude, as $B\propto r^{-(2+p)}$, with $p<1$ \cite{Thompson:2001ig}. 

Evaluating the power emitted per unit solid angle by analogy with Sect. \ref{II} would allow one to estimate the axion-induced flux density on Earth, S$_{\nu}=d\pazocal{P}/d\Omega/d^2/\Delta\nu$. In any case, performing numerical computations with the model outlined in \ref{Ap3}, more sophisticated but heavier than the 1D model developed in Sec. \ref{II}, is beyond the scope of this article, and is therefore left for a future work.

%\bibliographystyle{elsarticle-num}
\bibliographystyle{apsrev4-1}
\bibliography{sample631}



%% For example:

%% \bibitem[Aladro et al.(2015)]{Aladro15} Aladro, R., Martín, S., Riquelme, D., et al. 2015, \aas, 579, A101

%%\end{thebibliography}

\end{document}